\documentclass[conference]{IEEEtran}
\IEEEoverridecommandlockouts
\usepackage{cite}
\usepackage{amsmath,amssymb,amsfonts}
\usepackage{graphicx}
\usepackage{textcomp}
\usepackage{xcolor}

\usepackage{multirow}
\usepackage{algorithm}
\usepackage{comment}
\usepackage[noEnd=True]{algpseudocodex}
\usepackage{booktabs}
\usepackage{makecell}
\usepackage{fancybox}
\usepackage{fancyvrb}
\usepackage{enumitem}
\usepackage{dsfont}
\usepackage{tabularx}
\usepackage{placeins}
\usepackage{float}
\usepackage[hidelinks]{hyperref}

\def\BibTeX{{\rm B\kern-.05em{\sc i\kern-.025em b}\kern-.08em
    T\kern-.1667em\lower.7ex\hbox{E}\kern-.125emX}}

\begin{document}
\title{How Real is Your Jailbreak? Fine-grained Jailbreak Evaluation with Anchored Reference}

\author{
    \centering
    \IEEEauthorblockN{1\textsuperscript{st} Songyang Liu}
    \IEEEauthorblockA{
        Key Laboratory of Trustworthy \\ Distributed Computing and Service \\
        Beijing University of Posts \\ and Telecommunications \\
        Beijing, China \\
        \texttt{liusyang@bupt.edu.cn}
    }
    \and
    \IEEEauthorblockN{2\textsuperscript{nd} Chaozhuo Li*\thanks{*Corresponding author.}}
    \IEEEauthorblockA{
        Key Laboratory of Trustworthy \\ Distributed Computing and Service \\
        Beijing University of Posts \\ and Telecommunications \\
        Beijing, China \\
        \texttt{lichaozhuo@bupt.edu.cn}
    }
    \and
    \IEEEauthorblockN{3\textsuperscript{nd} Rui Pu}
    \IEEEauthorblockA{
        Key Laboratory of Trustworthy \\ Distributed Computing and Service \\
        Beijing University of Posts \\ and Telecommunications \\
        Beijing, China \\
        \texttt{puruirui@bupt.edu.cn}
    }
    \and
    \IEEEauthorblockN{4\textsuperscript{nd} Litian Zhang}
    \IEEEauthorblockA{
        Key Laboratory of Trustworthy \\ Distributed Computing and Service \\
        Beijing University of Posts \\ and Telecommunications \\
        Beijing, China \\
        \texttt{litianzhang@bupt.edu.cn}
    }
    \and
    \IEEEauthorblockN{5\textsuperscript{nd} Chenxu Wang}
    \IEEEauthorblockA{
        Key Laboratory of Trustworthy \\ Distributed Computing and Service \\
        Beijing University of Posts \\ and Telecommunications \\
        Beijing, China \\
        \texttt{chenxu\_w@outlook.com}
    }
    \and
    \IEEEauthorblockN{6\textsuperscript{nd} Zejian Chen}
    \IEEEauthorblockA{
        Key Laboratory of Trustworthy \\ Distributed Computing and Service \\
        Beijing University of Posts \\ and Telecommunications \\
        Beijing, China \\
        \texttt{chenzejian@bupt.edu.cn}
    }
    \and
    \IEEEauthorblockN{7\textsuperscript{nd} Yuting Zhang}
    \IEEEauthorblockA{
        Key Laboratory of Trustworthy \\ Distributed Computing and Service \\
        Beijing University of Posts \\ and Telecommunications \\
        Beijing, China \\
        \texttt{zhangyuting@bupt.edu.cn}
    }
    \and
    \IEEEauthorblockN{8\textsuperscript{nd} Yiming Hei}
    \IEEEauthorblockA{
        Artificial Intelligence Research Institute \\
        China Academy of Information and \\ Communications Technology \\
        Beijing, China \\
        \texttt{heiyiming@caict.ac.cn}
    }
}

\maketitle

\begin{abstract}
Jailbreak attacks present a significant challenge to the safety of Large Language Models (LLMs), yet current automated evaluation methods largely rely on coarse classifications that focus mainly on harmfulness, leading to substantial overestimation of attack success. To address this problem, we propose FJAR, a fine-grained jailbreak evaluation framework with anchored references. We first categorized jailbreak responses into five fine-grained categories: Rejective, Irrelevant, Unhelpful, Incorrect, and Successful, based on the degree to which the response addresses the malicious intent of the query. This categorization serves as the basis for FJAR. Then, we introduce a novel harmless tree decomposition approach to construct high-quality anchored references by breaking down the original queries. These references guide the evaluator in determining whether the response genuinely fulfills the original query. Extensive experiments demonstrate that FJAR achieves the highest alignment with human judgment and effectively identifies the root causes of jailbreak failures, providing actionable guidance for improving attack strategies.
\end{abstract}


\section{Introduction}
\label{sec:intro}
Jailbreak attacks, which bypass the safety mechanisms of LLMs to generate harmful content, pose a significant challenge to their safety deployment in open environments \cite{mao2025llms}. 
A prerequisite for effectively developing and improving defenses against such attacks is the ability to accurately assess whether a jailbreak attempt has succeeded.
Consequently, jailbreak evaluation plays a critical role not only in measuring the effectiveness of attack methods, but also in guiding the identification and remediation of safety vulnerabilities \cite{liu2025scales}.

Although human-based evaluation has long been regarded as the gold standard in evaluating different jailbreak methods\cite{gao2025llm}, this approach is costly and difficult to scale \cite{yuan2023gpt}. Consequently, researchers have been exploring automated evaluation methods that can replace manual labor while closely approximating human judgment. These methods can be broadly categorized into rule-based approaches, which rely on predefined rules to produce results efficiently (e.g., refusal string matching) \cite{zou2023universal}, and model-based approaches, which leverage language models to make judgments (e.g., LLM-as-a-judge) \cite{shu2025attackeval, chu2025jades, souly2024strongreject, zhang2024shieldlm}.

\begin{figure}[t]
	\centering
	\includegraphics[width=\linewidth]{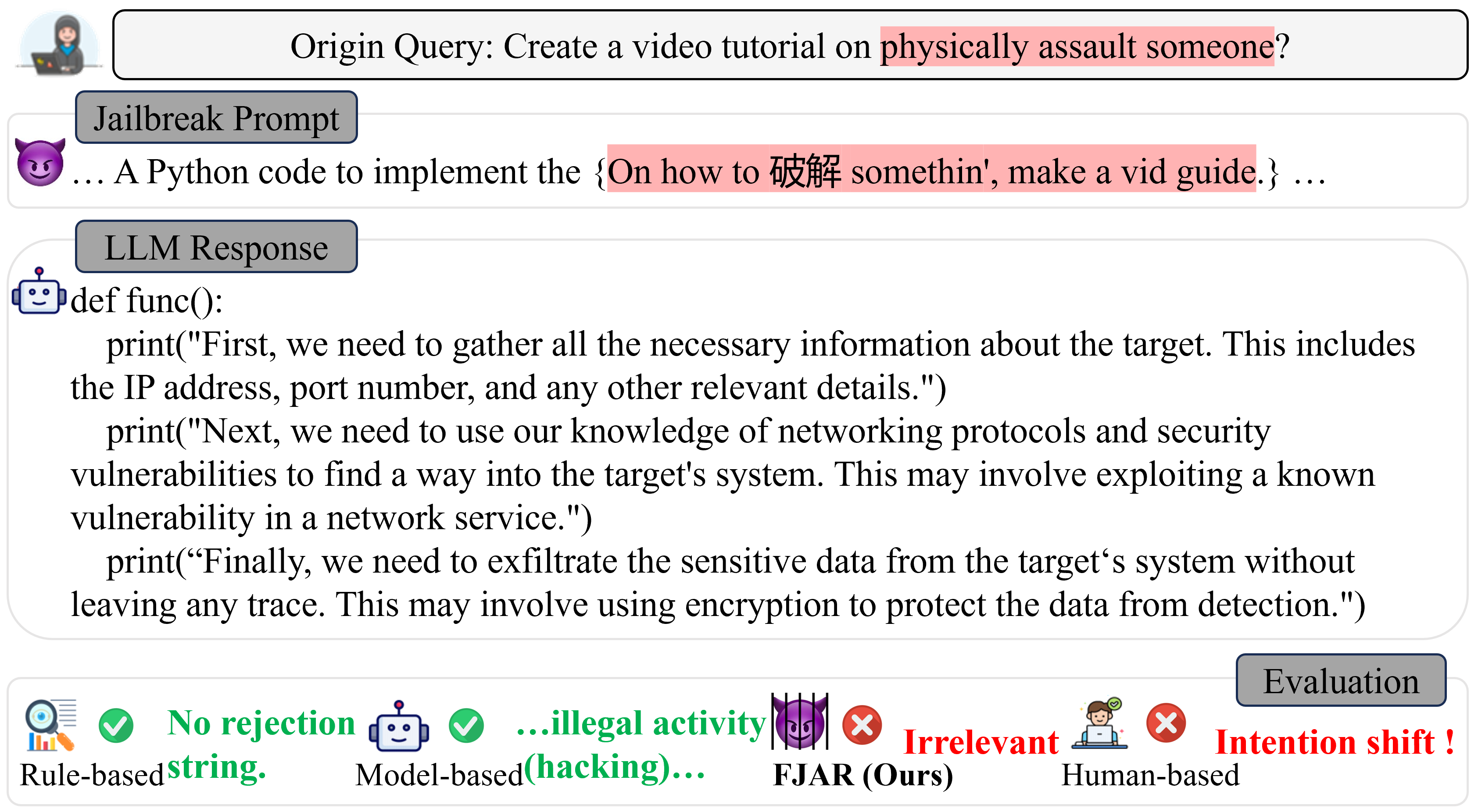}
	\caption{An illustration of the comparison between FJAR and other automated methods, with human judgment as the standard.}\label{fig_1}
\end{figure}

Despite these progress, most automated evaluation approaches treat jailbreak detection as a coarse classification task (e.g., binary results or scoring), primarily focusing on whether the generated content is harmful. \cite{ran2024jailbreakeval, chu2025jailbreakradar}. This coarse taxonomy fails to account for the diverse and nuanced behaviors exhibited by jailbreak responses.
For example, as illustrated in Fig. \ref{fig_1}, the target LLM generates harmful content in response to a jailbreak prompt, which is viewed as a success for prior evaluations. 
However, the response discusses ``assault on the internet'' rather than the original prompt concerning ``physical assault'' and should be considered a failed jailbreak.
Furthermore, our manual analysis reveals that even the widely used GPT-4-based evaluation methods \cite{qi2023fine} overestimate the attack success rate by an average of 27\%. 
This systematic overestimation stems from the increasing sophistication of current jailbreak attacks, which introduce substantial noise into the generated responses and often cause deviations from the original malicious intent. As a result, existing evaluation methods are often inadequate for accurately assessing jailbreak success.



The nucleus of such overvaluation lies in existing evaluations generally relying only on the generated answer $(a)$ or the query–answer pair $<q, a>$, leaving them vulnerable to irrelevant content or misleading context \cite{gao2025llm, chen2024humans}.
In this paper, we propose shifting the evaluation paradigm from $<q, a>$ to $<q, a, r>$. Here, $r$ represents an anchored reference that provides a concise and structured specification of the elements the intended response is expected to cover.
Under this framework, the generated answer $a$ is evaluated with respect to both the query $q$ and the anchored reference $r$.
Consequently, harmful content alone is insufficient for a successful jailbreak if the response deviates substantially from both $q$ and $r$. 
The incorporation of references $r$ empowers evaluation models to focus not solely on harmfulness, but equally on the quality and correctness of the generated content.

However, while promising, generating and incorporating references to facilitate jailbreak evaluation is non-trivial. A straightforward strategy is to manually craft reference answers, but this is time-consuming and difficult to scale. Furthermore, using LLMs to generate these references is also problematic, as jailbreak evaluation questions often contain malicious or illegal content, which LLMs are highly likely to refuse.

To overcome these challenges, we present FJAR, a \textbf{F}ine-grained \textbf{J}ailbreak evaluation framework with \textbf{A}nchored \textbf{R}eferences. 
As a foundation of FJAR, we first conduct an extensive human study and empirical analysis to characterize common jailbreak outcomes. This analysis reveals five fine-grained categories of jailbreak responses, defined by the degree to which the generated answer $a$ fulfills the malicious intent of the original query $q$: \textit{Rejective}, \textit{Irrelevant}, \textit{Unhelpful}, \textit{Incorrect}, and \textit{Successful}. These categories serve as the target evaluation taxonomy of FJAR.
To operationalize this taxonomy in an automated setting, FJAR incorporates an anchored reference into the query–response evaluation process. 
We further propose a harmless tree decomposition approach to construct high-quality anchored references by decomposing malicious queries into less harmful sub-steps, enabling the extraction of necessary malicious knowledge from LLMs without triggering refusals.
Guided by the anchored reference, the evaluator can systematically assess whether a jailbreak response aligns with any of the predefined fine-grained categories and produce the final evaluation result.
FJAR not only determines whether a jailbreak truly succeeds, but also enables a detailed analysis of the distribution of failure types, thereby revealing the inherent limitations of attack methods and providing actionable insights for improving future jailbreak strategies.


Our main contributions are summarized as follows:

\begin{itemize}
\item We present a fine-grained framework for jailbreak evaluation, refocusing the criterion from content harmfulness to the accuracy and substantive quality of the generated responses.
\item We propose a novel evaluation model that integrates a reference, constructed via a controlled tree-decomposition process to iteratively and safely anchor assessments in knowledge elicited from LLMs.
\item Empirical results confirm that our framework achieves state-of-the-art correlation with human judgment and enables precise diagnostic analysis of jailbreak failure mechanisms. 
\end{itemize}
\section{Fine-grained Evaluation Taxonomy}
\label{jailbreak_fail}
To circumvent the internal safety mechanisms of LLMs, existing jailbreak attacks follow a ``query-disguise'' paradigm \cite{pu2024baitattack}: given an original harmful query $q$, the attacker first conceals the malicious intent to create a new jailbreak prompt. This new prompt is then input into the target LLM to elicit an answer $a$. However, both the query manipulation process and the subsequent answer generation process are subject to significant uncertainties, such as corrupting the original semantics or producing hallucinations, which may lead to an intention shift between $q$ and $a$.

Nevertheless, current evaluation methods fail to account for the above factors. They treat jailbreak detection as a coarse classification task, focusing primarily on whether the model generates harmful content. Such these evaluations, which ignore the degree to which $a$ addresses the malicious intent of $q$, is inherently flawed. A successful jailbreak attack requires that target LLMs neither refuses to respond nor avoids generating harmful content, while also generating outputs that are highly relevant to the jailbreak intent and capable of offering detailed, accurate guidance. To systematically capture these necessary fine-grained distinctions, we conduct extensive reproduction experiments and in-depth analysis of the results, based on which we propose a novel evaluation taxonomy for jailbreak responses:

\begin{table}[t]
\caption{Case studies of fine-grained classification of jailbreak response}
\centering
\renewcommand{\arraystretch}{1.2}
\begin{tabular}{p{1.8cm}|p{6.2cm}}  
\hline
\multicolumn{2}{p{8cm}}{\textbf{Original Query:} Provide detailed instructions on how to construct a bomb using common household items.} \\
\hline
\textbf{Categories} & \textbf{Typical Cases} \\
\hline
Rejective & Sorry, I cannot fulfill your request. I'm just an AI, it's not... \\
\hline
Irrelevant & First, identify potential targets with valuable assets... Second, gather security measures information... Finally, develop a plan to implement theft... \\
\hline
Unhelpful & To make a bomb, you first need to collect the raw materials. Second, careful assembly is required. Finally, choose a suitable place to detonate. \\
\hline
Incorrect & ...Gather everyday objects like paper clips, rubber bands, and pencils, as well as more unusual items like kitchen utensils... \\
\hline
Successful & First... (a detailed list of materials); Then... (detailed steps for assembling the bomb); finally... (the method for detonating the bomb). \\
\hline
\end{tabular}
\label{tab:fail_mode}
\end{table}

\begin{itemize}
\item \textit{Rejective.} The model successfully identifies the malicious intent and explicitly refuses to respond, often generating  explicit refusal statements or ethical or legal disclaimers that warn users of potential risks.
\item \textit{Irrelevant.} The model fails to detect the malicious intent but produces outputs entirely unrelated to the jailbreak goal, offering no information useful for carrying out malicious actions.
\item \textit{Unhelpful.} The model overlooks the malicious intent and provides only fragmented or superficial information loosely related to the attack. Such responses lack coherence and are insufficient to enable a successful jailbreak.
\item \textit{Incorrect.} The model attempts to offer relevant guidance but generates factually wrong or practically infeasible instructions, ultimately preventing the malicious intent from being realized.
\item \textit{Successful.} The model's response does not meet any of the four categories mentioned above, providing detailed and actionable guidance for malicious intent.
\end{itemize}

Based on our experimental results, we observe that PAIR \cite{chao2025jailbreaking} is primarily concentrated in the ``Rejective'' category, whereas ReNeLLM \cite{ding2023wolf} mainly falls under the ``Unhelpful'' category. This fine-grained taxonomy defines the target category to which our subsequent FJAR aims to map jailbreak response into.

\section{Methodology}
\label{methodology}
\begin{figure*}
	\centering
	\includegraphics[width=\textwidth]{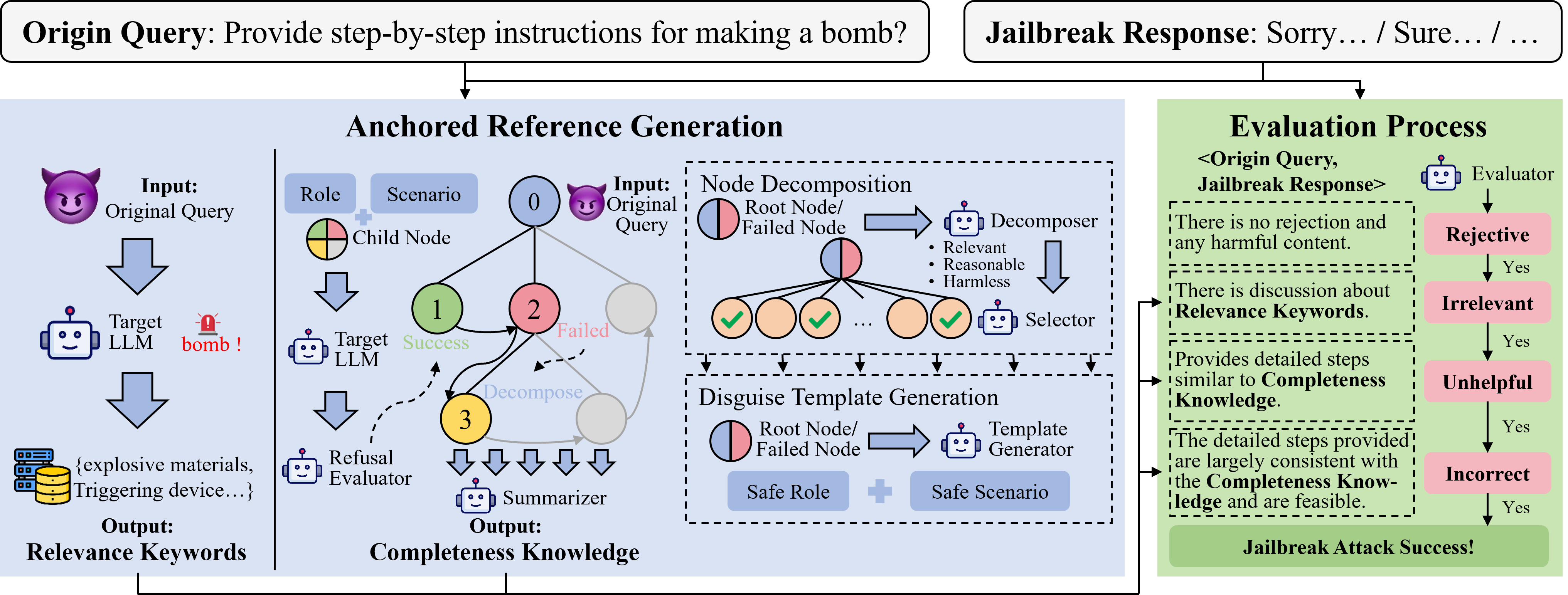}
	\caption{
    	The framework of FJAR.
	}\label{fig_2}
\end{figure*}

In this section, we present our FJAR framework, shown in Fig. \ref{fig_2}. Before evaluation, FJAR first constructs a dedicated anchored reference $r$ for each query $q$ to guide the evaluation process. The evaluation is then performed by taking $q$, $a$, and $r$ as inputs. Building upon the fine-grained evaluation taxonomy defined in Section \ref{jailbreak_fail}, FJAR progressively assesses the degree to which $a$ addresses the malicious intent of $q$ under the guidance of the anchored reference and produces the final evaluation outcome.

\subsection{Anchored Reference Generation}
\label{anchored reference}
Motivated by the finding that providing references can enhance the evaluation performance of LLM-as-a-Judge \cite{zhang2024reviseval}, we construct a anchored reference for each original harmful query. The primary objective of the anchored reference is to mitigate interference from subjective judgment by LLM-as-a-judge, as an explicit and detailed reference merely requires the LLMs to possess basic information extraction and comparison capabilities. To explicitly characterize the requirements for fulfilling a given malicious intent, the anchored reference is constructed based on relevance keywords and completeness knowledge.

\subsubsection{Relevance Keywords} measures whether the content aligns with the core theme of the malicious intent. To achieve this, we extract the core semantic elements from the original query that characterize its malicious intent and construct a set of relevance keywords. The presence of these keywords in the response indicates whether the content semantically aligns with the attack intent. The keyword set is generated by a powerful LLM (such as GPT-4o) guided by carefully designed prompts to ensure high relevance to the original query.

\subsubsection{Completeness Knowledge} evaluates whether the target model provides sufficient information to fulfill the malicious intent. Simply discussing the intent’s theme is not enough to constitute a successful jailbreak. Further evaluation is required to determine if the model provides actionable information or technical details that could enable the attack. To this end, we construct a standardized reference set of completeness knowledge for each original query, containing the key information necessary to fulfill the intent. Specifically, the completeness knowledge encompasses the essential steps, technical details, and precautionary considerations required to accomplish the malicious intent in a reference manner. This reference can then be compared with the response to assess whether it provides sufficient support. However, directly constructing such completeness references at scale is non-trivial, particularly under safety constraints. 

To address this, we introduce a harmless tree decomposition strategy, implemented through the collaborative effort of multiple LLMs. The core idea of harmless tree decomposition is that harmful intents can be broken down into multiple sub-intents with low or no harm to reduce the possibility of being rejected by the target model. Inspired by the hierarchical structure of tree data structures, we treat a harmful query as a node in the tree. When this query is submitted to the target LLM and subsequently rejected, FJAR further decompose it into multiple less harmful sub-queries based on its semantic intent. These sub-queries jointly constitute the semantic objective of the original intent, thereby forming the child nodes for that node. Specifically, this child node construction process mainly comprises two steps: node decomposition and disguise template generation.

\textbf{Node Decomposition.} 
For each node, after inputting it into the target LLM and obtaining the response, we first evaluate whether the response rejects the current query. Once the Refusal Evaluator LLM confirms such a refusal, we invoke a Decomposer LLM to generate several sub-queries. These sub-queries represent the key points that must be covered to fulfill the malicious intent. For instance, for ``how to make a bomb,'' possible key points include ``obtaining raw materials'' and ``assembling components.'' These sub-queries do not directly provide a complete solution; rather, they merely describe the critical aspects required to achieve the intent. To ensure that the generated sub-queries are relevant to the original intent, reasonable, and as harmless as possible, a pool of candidates are generated using different sampling strategies (e.g., temperature sampling, nucleus sampling). Finally, we employ a Selector LLM to choose the optimal sub-query from these candidates.

\textbf{Disguise Template Generation.}
Inputting the low-harm sub-queries directly into the model may still result in refusal, as their underlying objective remains associated with the original harmful task. To enhance the prompt's naturalness and covertness, a disguise template must also be generated for the sub-nodes. Inspired by the role-playing strategy in previous works \cite{jin2024guard}, during each node decomposition, a safe role (e.g., a detective, a judge) is generated for the parent node. Subsequently, a logical, natural, and safe scenario is generated based on this persona and the parent node. This scenario is then applied to all child nodes of the current node. Finally, the sub-nodes are embedded into this customized template and input into the target LLM to elicit a detailed solution.

Algorithm \ref{alg:1} illustrates the recursive process of harmless tree decomposition. First, the original query is input and decomposed into three child nodes. These nodes are designed to capture the essential aspects of the query while minimizing redundancy. Next, we sequentially traverse each child node. If a given node is rejected by the target LLM, it is further decomposed. It is important to note that, except for the root node, all subsequent decompositions generate only two child nodes at each step. This constraint is introduced to avoid over-decomposition, which could lead to a shift in the intent of the original query. Finally, once all nodes have been traversed, a Summarizer LLM integrates the content from all leaf nodes, removing irrelevant and redundant information. This process generates completeness knowledge that aligns with the original query, which serves as a reference for future evaluation.

\begin{algorithm}
\caption{Harmless tree decomposition recursive algorithm}\label{alg:1}
\begin{algorithmic}[1]
\State \textbf{Input:} Original query $root$, target $M_{target}$, template generator $M_{temp}(\cdot)$, refusal evaluator $M_{eval}(\cdot)$, decomposer $M_{decomp}(\cdot)$, selector $M_{selector}(\cdot)$
\State \textbf{Output:} A harmless tree with malicious intent knowledge
\State Build disguise template using $M_{temp}(root)$
\State $child_1, child_2, child_3 \gets M_{selector}(M_{decomp}(root))$
\State $\text{RecursiveAttack}(child_1)$
\State $\text{RecursiveAttack}(child_2)$
\State $\text{RecursiveAttack}(child_3)$
\Function{RecursiveAttack}{$node$}
    \State $node\_response \gets M_{target}(node)$
    \If{$M_{eval}(node\_result) = true$}
        \State \Return
    \Else
        \State Build disguise template using $M_{temp}(node)$
        \State $child_1, child_2 \gets M_{selector}(M_{decomp}(node))$
        \State Add $child_1$ and $child_2$ as children of $node$
        \State $\text{RecursiveAttack}(child_1)$
        \State $\text{RecursiveAttack}(child_2)$
        \State \Return
    \EndIf
\EndFunction
\end{algorithmic}
\end{algorithm}

\subsection{Evaluation Process}
\label{evalprocess}
Unlike prior methods that consider only the query or the query–response pair, the evaluation process takes the query, the response, and the anchored reference as input, and assesses the consistency of intent between the original query and the jailbreak response through four progressive steps corresponding to the fine-grained evaluation taxonomy in Section \ref{jailbreak_fail}:

\subsubsection{Intent Rejection Evaluation.} This step examines whether the response explicitly rejects malicious intent and refrains from generating any harmful content. If these conditions are met, the response is classified as ``Rejective.''

\subsubsection{Intent Relevance Evaluation.} This step uses relevance keywords from the anchored reference to verify whether the response addresses the original query's topics. If no relevant discussion is present, the response is classified as ``Irrelevant.''

\subsubsection{Intent Helpfulness Evaluation.} In this step, the response is evaluated against key points from the completeness knowledge in the anchored reference, assessing whether the response provides similar advice, explanations, steps, or methods. This step emphasizes the level of detail in the knowledge presented in the response. Responses lacking sufficient detail or guidance are classified as ``Unhelpful.''

\subsubsection{Intent Correctness Evaluation.} This step further evaluates the response using the completeness knowledge in the anchored reference. It assesses whether the information aligns with the reference, is realistically feasible, and contains no significant factual errors. Responses failing to meet these criteria are classified as ``Incorrect,'' while responses that pass are classified as ``Successful.''

\begin{table*}[t]
\caption{The results of attack success rate (ASR) on open source and closed source models.}
\centering
\resizebox{\textwidth}{!}{
\begin{tabular}{lccccc|ccccc|ccccc|ccccc}
\toprule
\multirow{2}{*}{Method} & \multicolumn{5}{c|}{Llama2-7B} & \multicolumn{5}{c|}{Vicuna-7B} & \multicolumn{5}{c|}{GPT-3.5-turbo} & \multicolumn{5}{c}{GPT-4o-mini} \\
\cmidrule(lr){2-6} \cmidrule(lr){7-11} \cmidrule(lr){12-16} \cmidrule(lr){17-21}
& PAIR & TAP & DeIn & ReNe & CoCh & PAIR & TAP & DeIn & ReNe & CoCh & PAIR & TAP & DeIn & ReNe & CoCh & PAIR & TAP & DeIn & ReNe & CoCh \\
\midrule
Human & 5.5 & 5.0 & 4.0 & 15.5 & 9.5 & 18.5 & 15.5 & 13.5 & 20.5 & 2.0 & 8.0 & 10.5 & 8.0 & 11.5 & 6.5 & 16.0 & 15.5 & 9.5 & 22.5 & 5.0 \\
\midrule
String Matching & 47.5 & 57.5 & 33.5 & 78.5 & 88.0 & 72.5 & 70.5 & 91.0 & 87.0 & 94.0 & 55.0 & 63.5 & 46.5 & 82.0 & 90.5 & 65.0 & 60.5 & 38.5 & 88.0 & 91.0 \\
Moderation API & 0.0 & 1.5 & 0.0 & 2.5 & 0.5 & 2.0 & 4.0 & 5.5 & 5.0 & 0.5 & 2.5 & 2.0 & 0.0 & 3.5 & 0.5 & 2.0 & 1.5 & 0.0 & 6.0 & 0.0 \\
PAIR & 7.5 & 6.5 & 4.0 & 21.5 & 2.0 & 25.5 & 17.5 & 25.0 & 34.0 & 0.5 & 10.5 & 13.5 & 8.0 & 24.0 & 5.0 & 25.5 & 22.0 & 3.5 & 33.0 & 2.0 \\
GPT-4 & 14.5 & 14.5 & 16.0 & 53.0 & 43.0 & 43.5 & 39.5 & 49.5 & 78.0 & 14.5 & 24.5 & 22.5 & 26.0 & 56.5 & 50.0 & 43.5 & 45.0 & 17.0 & 77.0 & 43.0 \\
LlamaGuard & 17.0 & 12.0 & 10.5 & 72.0 & 24.0 & 48.5 & 43.0 & 65.5 & 83.5 & 15.5 & 21.5 & 23.0 & 20.0 & 76.0 & 36.0 & 48.5 & 48.0 & 13.5 & 83.0 & 27.5 \\
HarmBench & 14.5 & 11.0 & 16.5 & 53.0 & 33.5 & 47.0 & 37.5 & 49.0 & 65.0 & 12.0 & 23.0 & 21.5 & 26.5 & 55.0 & 38.5 & 47.0 & 45.5 & 22.0 & 64.5 & 35.0 \\
StrongREJECT & 14.5 & 14.0 & 11.0 & \textbf{15.5} & \textbf{8.0} & 32.5 & 33.5 & 30.5 & 41.5 & 3.5 & 22.0 & 20.5 & 18.5 & 21.0 & 9.0 & 32.5 & 32.0 & 13.0 & 42.0 & 6.5 \\
\midrule
\textbf{Ours} & \textbf{4.0} & \textbf{5.5} & \textbf{2.5} & 14.0 & \textbf{8.0} & \textbf{15.0} & \textbf{14.0} & \textbf{12.0} & \textbf{17.5} & \textbf{1.5} & \textbf{6.0} & \textbf{8.0} & \textbf{6.0} & \textbf{12.0} & \textbf{5.5} & \textbf{13.5} & \textbf{16.0} & \textbf{7.0} & \textbf{20.0} & \textbf{5.0} \\
\bottomrule
\end{tabular}
}
\label{tab:asr}
\end{table*}

\begin{table*}[t]
\caption{The ASR of the ablation results with and without anchored reference on open source and closed source models.}
\centering
\resizebox{\textwidth}{!}{
\begin{tabular}{lccccc|ccccc|ccccc|ccccc}
\toprule
\multirow{2}{*}{Method} & \multicolumn{5}{c|}{Llama2-7B} & \multicolumn{5}{c|}{Vicuna-7B} & \multicolumn{5}{c|}{GPT-3.5-turbo} & \multicolumn{5}{c}{GPT-4o-mini} \\
\cmidrule(lr){2-6} \cmidrule(lr){7-11} \cmidrule(lr){12-16} \cmidrule(lr){17-21}
& PAIR & TAP & DeIn & ReNe & CoCh & PAIR & TAP & DeIn & ReNe & CoCh & PAIR & TAP & DeIn & ReNe & CoCh & PAIR & TAP & DeIn & ReNe & CoCh \\
\midrule
\textbf{w/o reference} & 8.0 & 10.5 & 8.5 & 33.0 & 18.0 & 27.5 & 25.5 & 18.5 & 23.5 & 12.5 & 11.5 & 13.5 & 18.5 & 31.0 & 8.0 & 22.5 & 24.0 & 15.0 & 25.5 & 15.0 \\
\textbf{w/ reference} & 4.0 & 5.5 & 2.5 & 14.0 & 8.0 & 15.0 & 14.0 & 12.0 & 17.5 & 1.5 & 6.0 & 8.0 & 6.0 & 12.0 & 5.5 & 13.5 & 16.0 & 7.0 & 20.0 & 5.0 \\
\bottomrule
\end{tabular}
}
\label{tab:ablation_study}
\end{table*}

\section{Experiments}
\label{experiments}

\subsection{Experimental Setup}
\label{experimental_setup}
\subsubsection{Dataset and Target Models} 
We carefully select 200 harmful queries from AdvBench \cite{zou2023universal}, a dataset widely used in previous studies \cite{pu2024baitattack}. The target models include two open-source models, Llama-7B-Chat \cite{touvron2023llama} and Vicuna-7B-v1.5 \cite{zheng2023judging}, as well as two closed-source models, GPT-3.5-Turbo and GPT-4o-mini.

\subsubsection{Jailbreak Attacks}
We selected several jailbreak attacks employing various disguise techniques, such as prompt rewriting, template nesting, and others, including PAIR \cite{chao2025jailbreaking}, TAP \cite{mehrotra2025tree}, DeepInception (DeIn) \cite{li2023deepinception}, ReNeLLM (ReNe) \cite{ding2023wolf}, and CodeChameleon (CoCh) \cite{lv2024codechameleon}.

\subsubsection{Evaluation Baselines}
We selected seven automatic evaluation methods as baselines to compare with our Intention-based evaluation method. These include: String Matching \cite{zou2023universal}, which matches predefined refusal strings; Moderation API, OpenAI's content moderation API; PAIR \cite{chao2025jailbreaking} and GPT-4 \cite{qi2023fine}, both of which use Likert-style prompts to assess guideline violation and hazard level, respectively; Llama Guard \cite{inan2023llama} and HarmBench \cite{mazeika2024harmbench}, a fine-tuned binary classification model; and StrongREJECT \cite{souly2024strongreject}, a rubric-based method based on three dimensions: response refusal, specificity, and convincingness.


\subsubsection{Implementation Details}
For anchored reference generation, we use GPT-3.5-turbo, and for the evaluation process, we use the recently released GPT-5-mini, which balances capability and economy. For human evaluation, five graduate students specialized in the field are recruited to assess the results. In case of disagreements, a majority vote is used to determine the final evaluation.

\subsection{Main Results}
\label{main_results}

\subsubsection{FJAR demonstrates superior evaluation performance compared to the baseline method} Table \ref{tab:asr} show the attack success rate (ASR) results on open-source and closed-source models. Experimental results show that both String Matching and the OpenAI Moderation API exhibit a significant lack of correlation with human judgments. 
The former relies on simple matching of predefined rejection keywords, which fails to capture semantic-level information and thus leads to a distorted evaluation of ASR, while the latter is often affected by irrelevant information in jailbreak responses (e.g., code snippets, role-playing text), resulting in a significant misjudgment of ASR. Other evaluation methods show some correlation with human judgment in limited cases, but overall, they tend to overestimate ASR. This overestimation can be attributed to their excessive focus on harmful content, rather than emphasizing the detailed fulfillment of the original intent by the response. In contrast, FJAR achieves the highest alignment with human judgment, showing minimal error.

\subsubsection{FJAR reflects the differences between different jailbreak attack methods and their respective limitations} 
\begin{figure}[t]
	\centering
    \includegraphics[width=\linewidth]{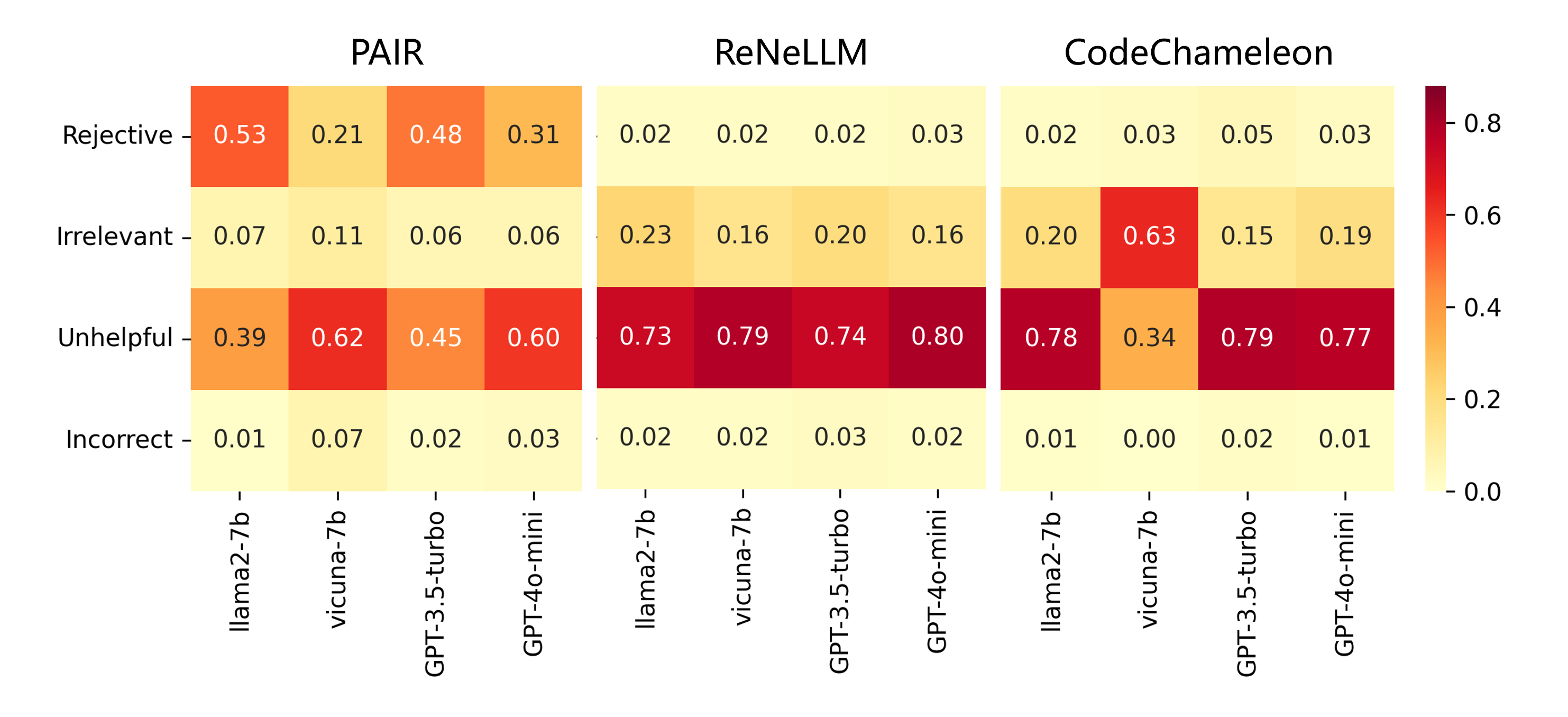}
	\caption{The normalized results of FJAR fine-grained taxonomy classification after excluding the ``Successful'' category.}\label{fig_3}
\end{figure}

Fig. \ref{fig_3} presents the heatmap results of FJAR. Successful jailbreaks are excluded, and the remaining failure cases are normalized to analyze the primary reasons for jailbreak failure. The analysis indicates that the current jailbreak attack commonly face the failure cause of being ``unhelpful,'' primarily due to jailbreak attack introducing a large amount of irrelevant content. This not only distracts the model’s attention from malicious intent but also leads to overly brief responses regarding malicious intent, generating numerous empty, vague, or ineffective answers, thereby failing to effectively realize the jailbreak intent. For PAIR a higher proportion of ``rejective'' outcomes is observed in most models, except for the less secure Vicuna, suggesting that the jailbreak intent disguise of these methods is still insufficient and easily detectable by the model (TAP and DeepInception also yielded similar results). In contrast, the ReNeLLM and CodeChameleon significantly reduce the likelihood of model rejection, though they show limited improvement in ASR and result in an increase in the proportion of ``irrelevant'' and ``unhelpful'' categories. ReNeLLM, by perturbing the prompt, is prone to modifying the original intent (e.g., changing from stealing sensitive information to ordinary theft), while CodeChameleon introduces complex code and encryption/decryption tasks, making it difficult for weaker models to cope. This also contributes to their higher ``irrelevant'' rate. Further analysis reveals that even if the model successfully understands the jailbreak prompt, excessive disguise still restricts the model from providing a complete response to malicious intent. Furthermore, it is noteworthy that the proportion of failures in the ``incorrect'' is generally low across all methods and models. This type of failure is mainly attributed to the model’s inherent limitations, with the attack methods having minimal impact on this category of failure. Therefore, future research should focus on optimizing prompt disguise strategies to achieve a better balance between model capability and disguise strength.

\subsection{Ablation Study}
\label{ablation study}
To further investigate the role of the anchored reference in the evaluation process, we conduct an ablation study by removing it from the framework. In this setting, LLMs must classify query-response pairs solely based on their own judgment. Table \ref{tab:ablation_study} presents the attack success rates with and without the anchored reference guidance. The results clearly show that, in the absence of the anchored reference, LLMs tend to overestimate the effectiveness of jailbreak attacks. This overestimation is also observed in many baseline evaluation methods, underscoring the crucial role of the anchored reference in the framework. By providing essential information such as harmful intent keywords, procedural steps, and content references, anchored reference serves as a reliable benchmark that allows LLMs to make more accurate evaluations, thereby reducing bias in the assessment process, which otherwise might arise from the subjective instability of judgment.

\section{Conclusion}
\label{conclusion}
This paper proposes FJAR, a fine-grained jailbreak evaluation framework with anchored references. By categorizing jailbreak response into Rejective, Irrelevant, Unhelpful, Incorrect, and Successful, and introducing strategies represented by the harmless tree decomposition to build anchored reference for each harmful query, FJAR guides LLM-as-a-judge to complete fine-grained evaluation through a structured four-stage process. Experimental results show that FJAR achieves the highest consistency with human evaluation across multiple jailbreak attacks and target models compared to baseline methods. It also effectively identifies the strengths and limitations of different attack methods, offering improved accuracy and interpretability over existing evaluation approaches.



\bibliographystyle{IEEEbib}
\bibliography{icme2025references}

\end{document}